\pdfoutput=1

\documentclass[12pt]{article}
\usepackage{graphicx}
\usepackage{cite}

\newcommand\pubnumber{DPF2015-267}
\newcommand\pubdate{\today}

\def\napoli{Southern Methodist University\\
Dallas, TX 75206, USA}

\def\Title#1{\begin{center} {\Large Drell-Yan Production of W/Z at the LHC\\
with Protons and Heavy Nuclei} \end{center}}
\def\Author#1{\begin{center}{ \sc D. Benjamin Clark} \end{center}}
\def\Address#1{\begin{center}{ \it Southern Methodist University\\
Dallas, TX 75206, USA } \end{center}}

\newcommand\pubblock{\rightline{\begin{tabular}{l} \pubnumber\\
         \pubdate  \end{tabular}}}
\newenvironment{Abstract}{\begin{quotation}  }{\end{quotation}}
\newenvironment{Presented}{\begin{quotation} \begin{center} 
             PRESENTED AT\end{center}\bigskip 
      \begin{center}\begin{large}}{\end{large}\end{center} \end{quotation}}

\def\beq{\begin{equation}}
\def\eeq#1{\label{#1}\end{equation}}
\def\eeqn{\end{equation}}

\def\beqa{\begin{eqnarray}}
\def\eeqa#1{\label{#1}\end{eqnarray}}
\def\eeqan{\end{eqnarray}}

\let\bar=\overbar

\def\Dslash{\not{\hbox{\kern-4pt $D$}}}
\def\dslash{\not{\hbox{\kern-2pt $\del$}}}

\def\msb{{\bar{\ssstyle M \kern -1pt S}}}

\begin{document}
\begin{titlepage}
\pubblock

\vfill
\Title{Drell-Yan Production of W/Z at the LHC\\
with Protons and Heavy Nuclei}
\vfill
\Author{ D. Benjamin Clark\support}
\Address{\napoli}
\vfill
\begin{Abstract}
Drell-Yan W/Z electroweak boson production at the LHC is an essential standard 
candle which is used for calibration of beam luminosity and detector 
properties. In addition to proton-proton collisions, the LHC has measured heavy 
nuclei lead-lead and proton-lead W/Z production. Inclusion of these data sets 
in future fits can provide discriminating information of the nuclear 
modifications present in the Parton Distribution Functions (PDFs). We present 
an ongoing analysis of W/Z production in lead-lead and proton-lead collisions 
at the LHC using the \texttt{nCTEQ15} nuclear Parton Distribution Functions 
(nPDFs) including uncertainties. The cross sections are calculated at 
NLO with the \texttt{FEWZ\ 2.1} program at 2.76 and 5.02 TeV. We identify 
promising observables for the observation of the nuclear modifications.
\end{Abstract}
\vfill
\begin{Presented}
DPF 2015\\
The Meeting of the American Physical Society\\
Division of Particles and Fields\\
Ann Arbor, Michigan, August 4--8, 2015\\
\end{Presented}
\vfill
\end{titlepage}
\def\thefootnote{\fnsymbol{footnote}}
\setcounter{footnote}{0}
%

\section{Introduction}

Parton Distribution Functions (PDFs) are an essential component for the 
calculation of any observable in hadronic collisions. Collisions involving 
Heavy Nuclei (HI) require input from nuclear PDFs (nPDFs) which show 
significant modifications to free proton PDFs. These modifications were first 
seen by the EMC collaboration in 1983~\cite{Aubert:1983xm} and later 
measured and described by SLAC in the 1980s and '90s (e.g. see~\cite{Gomez:1993ri}). 
The nuclear modifications across the $x$ range are generally 
known as shadowing, anti-shadowing, the EMC effect, and Fermi motion. 

Recently, the \texttt{nCTEQ} collaboration has produced a new set of nPDFs for 
19 different values of $A$ as explained in~\cite{Kovarik:2015cma,
Schienbein:2009kk}. These nPDFs can be used 
to compute observables for HI collisions at NLO along with Hessian estimations of 
the PDF errors. The nuclear modifications to the free proton PDF are described 
by and $A$-dependent parameterization fit to data from DIS, D-Y, and 
inclusive pion production. The $A$-dependence is given to the parameters 
of the free proton PDF such that a value of $A,Z = 1$ will reproduce the free 
proton. 

\begin{figure}[h]
\centering
\includegraphics[width=0.45\textwidth]{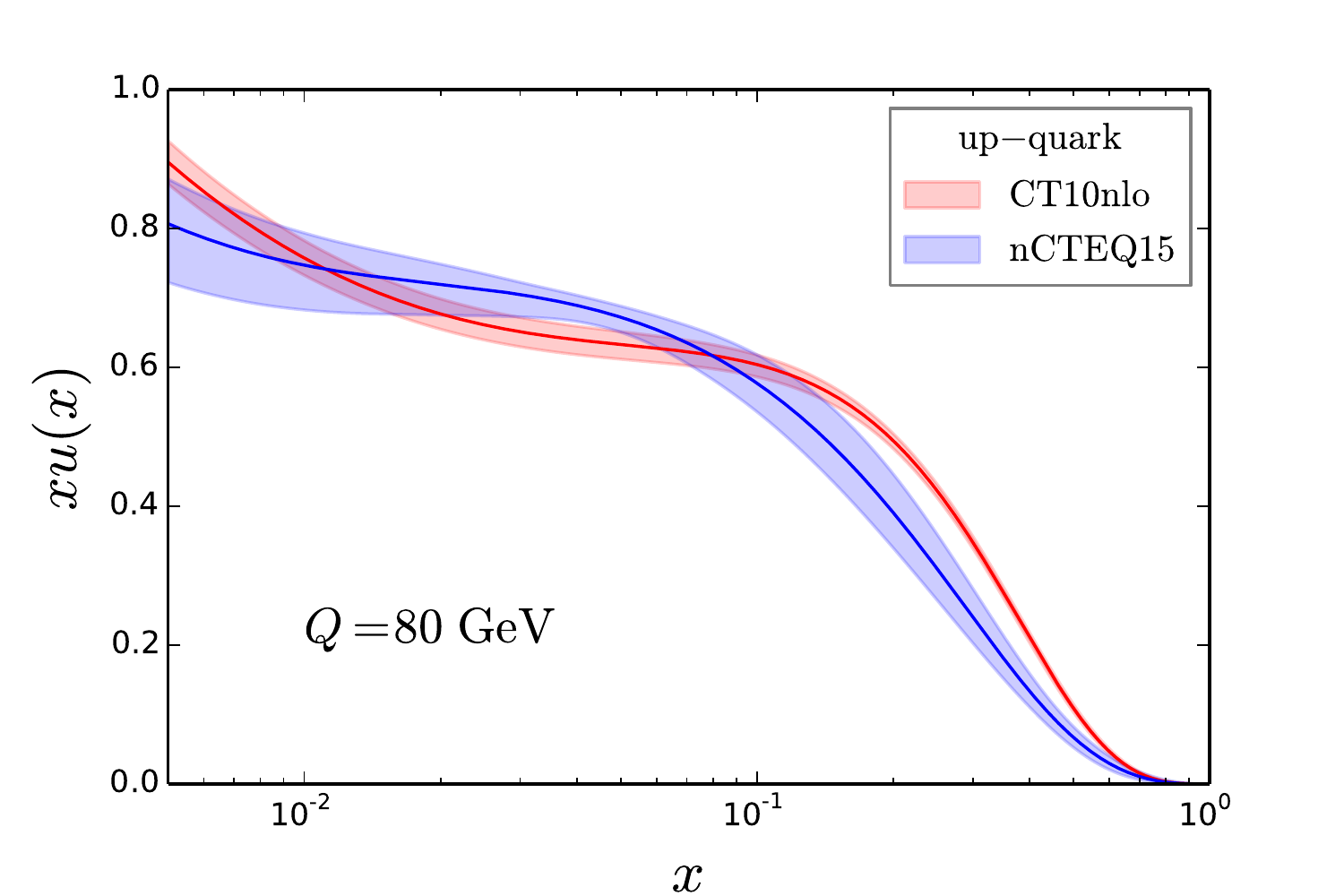}
\includegraphics[width=0.45\textwidth]{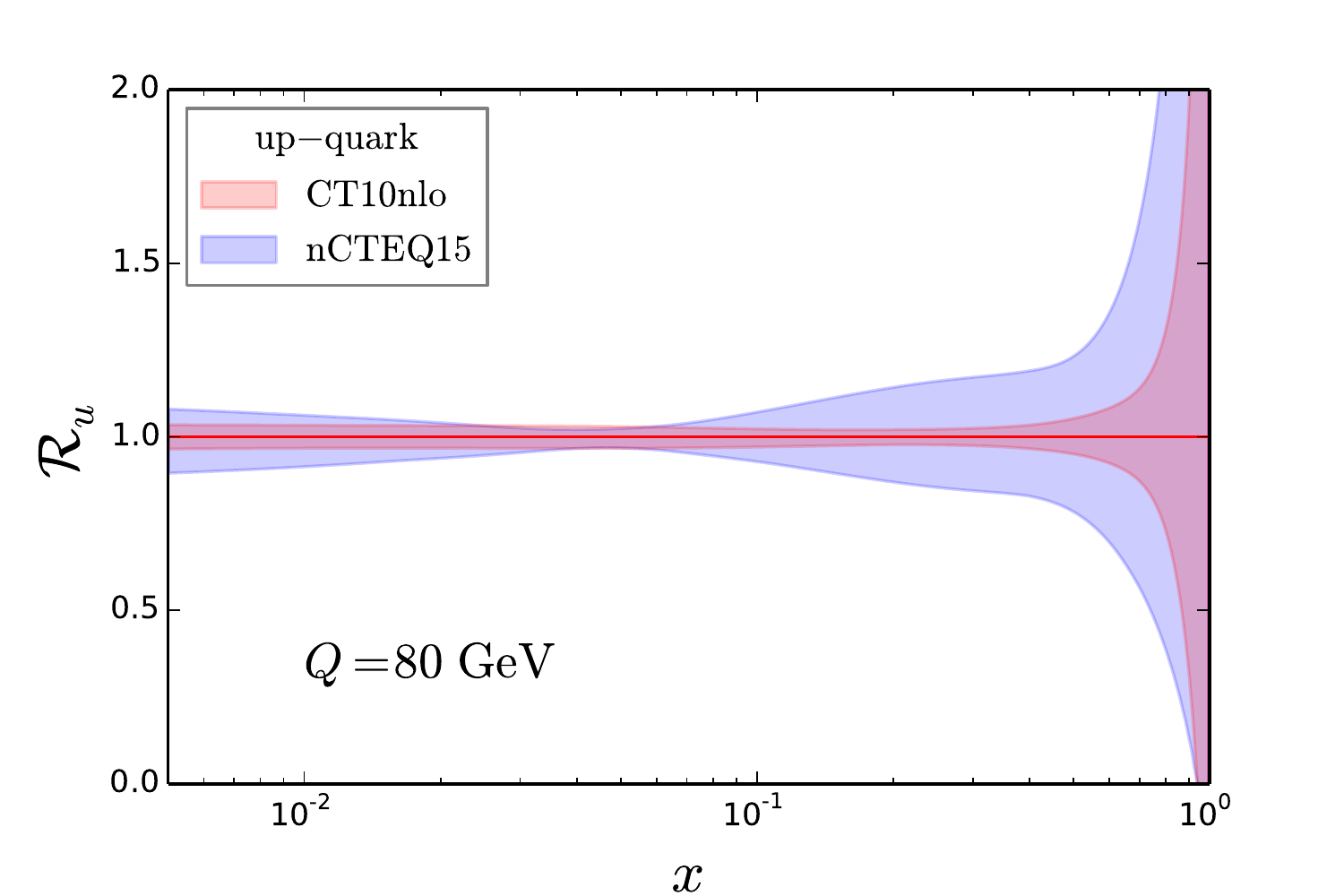}
\caption{The left plot shows the \texttt{nCTEQ15} bound Pb PDF compared to 
the CT10 free proton. The initial state modifications are apparent in the 
softening of the valance quark peak. On the right the ratio demonstrates 
that the free proton PDF is much more tightly constrained over the entire 
range of $x$.}
\label{fig:up_pdf}
\end{figure}

Hessian error sets are provided for the errors on the nuclear parameters only. 
These errors are much larger than the errors on the proton PDF parameterization 
as can be seen in Fig.~\ref{fig:up_pdf}. Therefore errors shown in this note are due to the 
nuclear parameters only. At this time, no statistically sound method to combine 
the errors on the free proton parameters with the errors on the nuclear 
parameters exists in the literature.

The full nuclear target PDF is constructed from the bound neutron and proton 
PDFs, $f_i^{n/A} (x,Q)$ and $f_i^{p/A} (x,Q)$ respectively as,
\begin{equation}
  f_{i}^{(Q,Z)}(x,Q) = \frac{Z}{A}\ f_{i}^{p/A}(x,Q) + \frac{(A-Z)}{A}\ 
  f_{i}^{n/A}(x,Q)\, ,\label{eq:pdf}
\end{equation}
where $A-Z$ is the number of neutrons in the nuclear target. In the above equation 
the bound neutron PDF is found by the approximate isospin symmetry present at NLO.
That is, the neutron PDF is constructed by the replacements 
$u \longleftrightarrow d$ and $\bar{u} \longleftrightarrow \bar{d}$ in the proton PDF. 

\section{Vector Boson Production}

The LHC produces many electroweak bosons at high rapidity and properties of these 
bosons have been well-studied~\cite{Chatrchyan:2014csa,Aaij:2014pvu,Khachatryan:2015hha,
Aad:2012ew,Aad:2014bha,Chatrchyan:2012nt,Aad:2010yt,Aad:2015gta}. In the 
factorization formalism, the cross section for lepton pair production is 
written as a convolution of the partonic cross section with the PDFs for each 
of the colliding hadrons as,
\begin{equation}
  \frac{d\sigma}{dQ^2\,dy} = \sum_{a,b}\int_0^1 d\xi_1 \int_0^1 d\xi_2\frac{d\hat{\sigma}^{a,b}}{dQ^2dy}f_{a/A}(\xi_1)
  f_{b/B}(\xi_2)\, .\label{eq:xsec}
\end{equation}

The PDFs are integrated over the longitudinal momentum fractions $\xi_{1,2}$ of 
the partons involved in the interaction. In the collinear approximation, the 
longitudinal momentum fraction becomes the Bjorken scaling variable $x$. For 
fixed partonic energy $Q$, a change in rapidity, $y$, corresponds to a change in the 
relative fractions of $x_{1,2}$ for the incoming partons. Thus, measurement of 
the rapidity distribution of lepton pairs produced in hadronic collisions gives 
a handle on the PDF distributions in $x$.

In addition, comparison of different observables involving electroweak bosons allows for 
flavor decomposition in the measured PDFs. This is because different quark 
flavors contribute to the production of different vector bosons. For example, 
the major contribution to the $W^+$ cross section is the $u - \bar{d}$ 
interaction while the major contribution to 
the $W^-$ cross section is the $d - \bar{u}$ interaction. These two measurements are 
very sensitive to the $u$ and $d$ parameterization in the fits and allow for 
comparison of the various nPDF releases. 

\begin{figure}[t]
\centering
\includegraphics[width=0.45\textwidth]{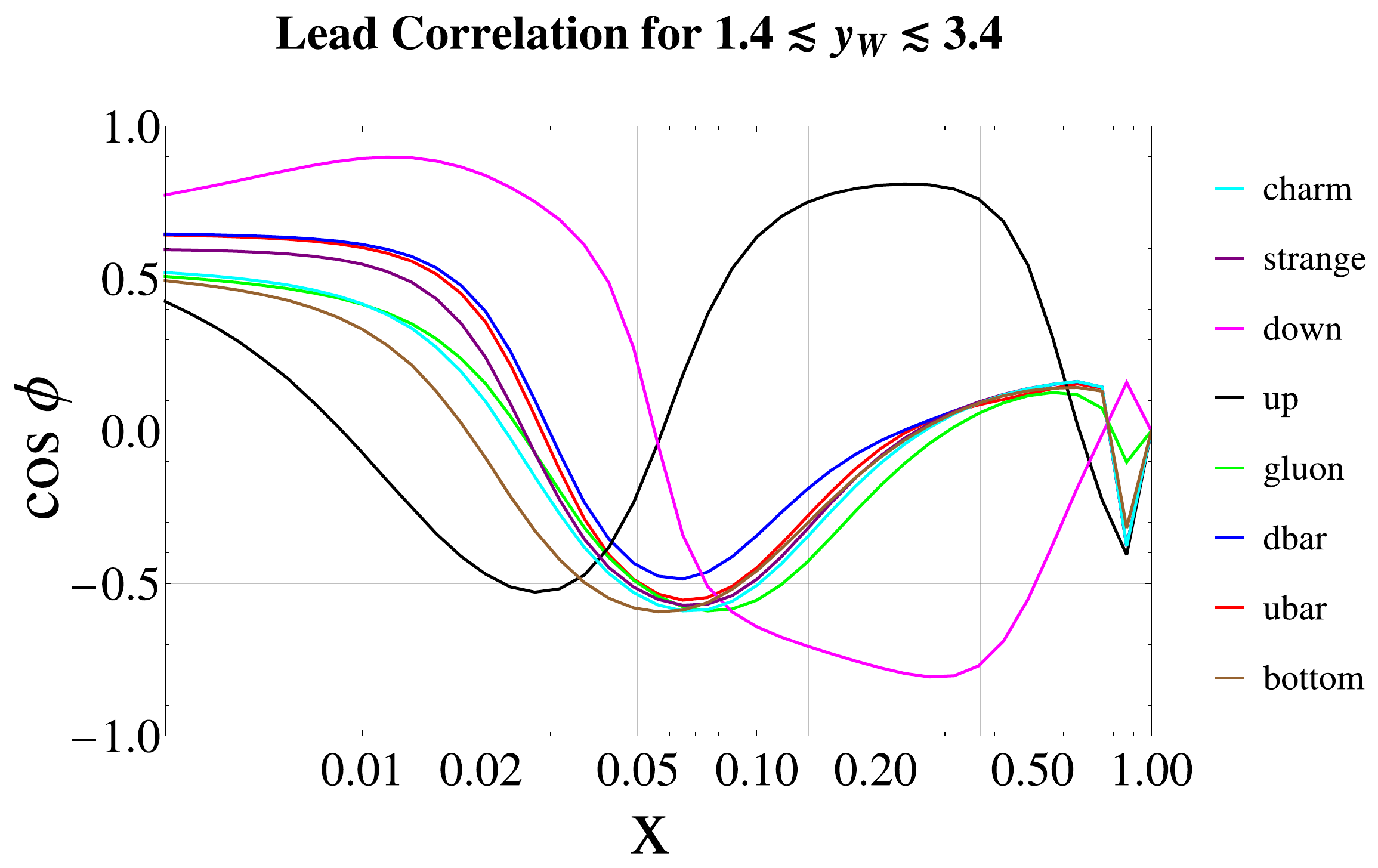}
\includegraphics[width=0.45\textwidth]{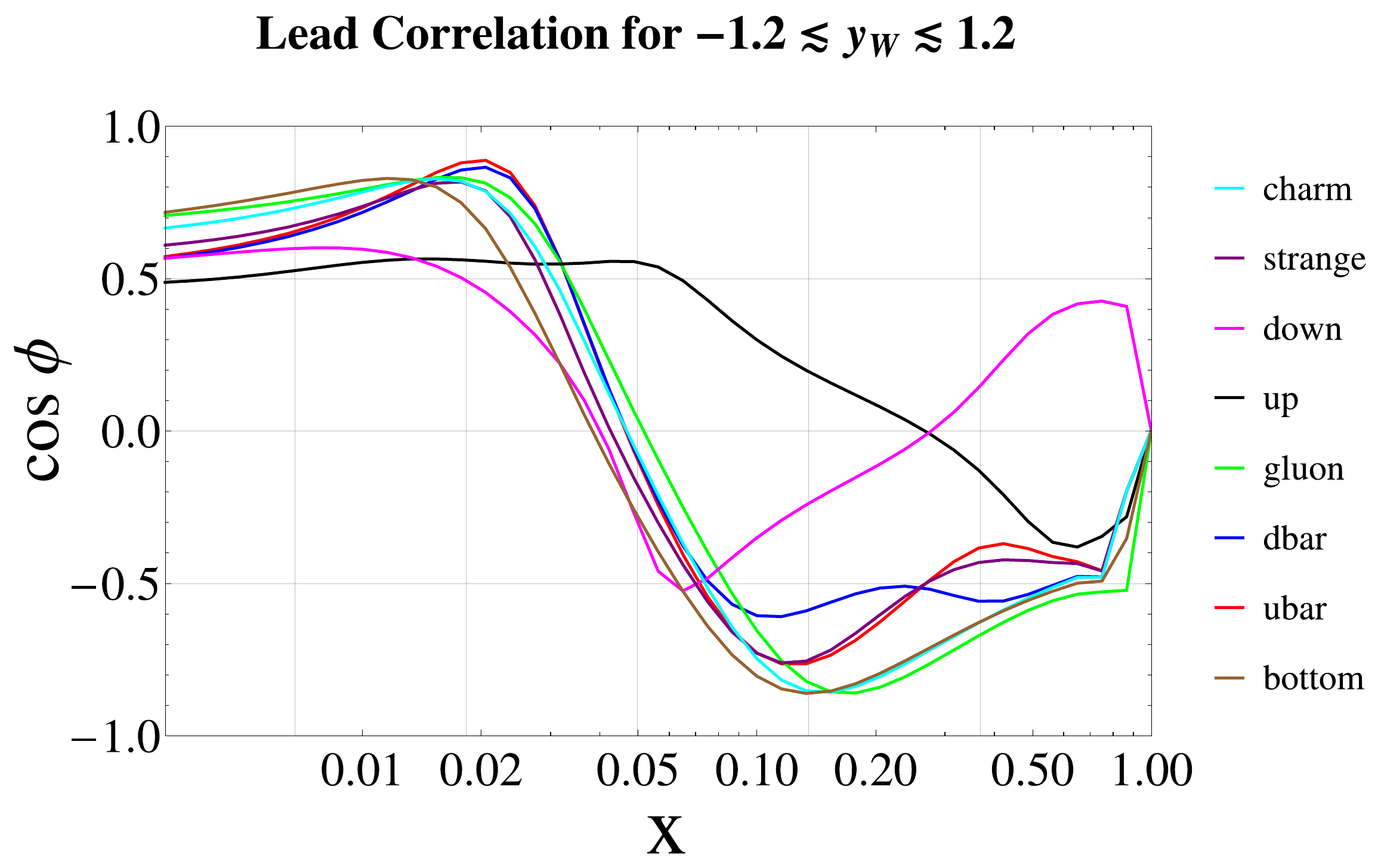}
\caption{The cosine of the correlation angle for the $W^+$ cross section for 
the various flavors as a function of $x$. The left plot shows the 
high rapidity bins  and the right plot shows the central region.}
\label{fig:cor_angle}
\end{figure}

The contribution of the error on each flavor to the overall error on the 
measurements can be estimated by looking at the cosine of the correlation angle 
as defined in~\cite{Nadolsky:2008zw}, see Fig.~\ref{fig:cor_angle}. A correlation of 
$\approx 1$ indicates that the error on the measurement of this observable is being 
driven by the error on that quark flavor. Furthermore, a high correlation indicates 
that the inclusion of this measurement in the fit will directly constrain the error 
on that quark flavor. The correlation presented here also shows that the $u$ and $d$ quark 
PDFs are anti-correlated over the range of $x$. This anti-correlation results in the  
decomposition of these flavors and helps to constrain their PDFs.

\section{Results}

All calculations in this preliminary study have been preformed at NLO with 
\texttt{FEWZ\ 2.1}~\cite{Gavin:2012sy} modified to accept two different PDFs for the two 
interacting nuclei. The code was benchmarked against the original version of 
\texttt{FEWZ\ 2.1} for the pp and PbPb cases and demonstrated exact agreement. 
All calculations are preformed and presented in the center of momentum (COM) 
frame. The COM frame for PbPb collisions corresponds to the lab frame as the 
two colliding beams have the same energy. For the pPb collisions the lab frame 
is shifted relative to the COM frame by $\eta_{LAB} = \eta_{COM} + 0.465$. This 
rapidity shift is not accounted for in this study and will be 
included when comparing to the available data.

\subsection{Lead Lead Collisions}

The cross sections for PbPb collisions at 2760 GeV COM energy at the LHC were 
computed and compared to predictions for a lead nucleus constructed from CT10 
free proton PDFs according to Eq.~(\ref{eq:pdf}).  While the effects of the 
initial state modifications are 
visible in the cross sections, these effects are washed out 
by the isoscalarity of the Pb nucleus; that is, the uncertainty bands for the 
two calculations overlap significantly. This fact coupled with a systematically 
dominated measurement leaves little hope for detecting the nuclear modifications 
to electroweak boson production in PbPb collisions at this energy.

\subsection{Proton Lead Collisions}

\begin{figure}[t]
\centering
\includegraphics[width=0.45\textwidth]{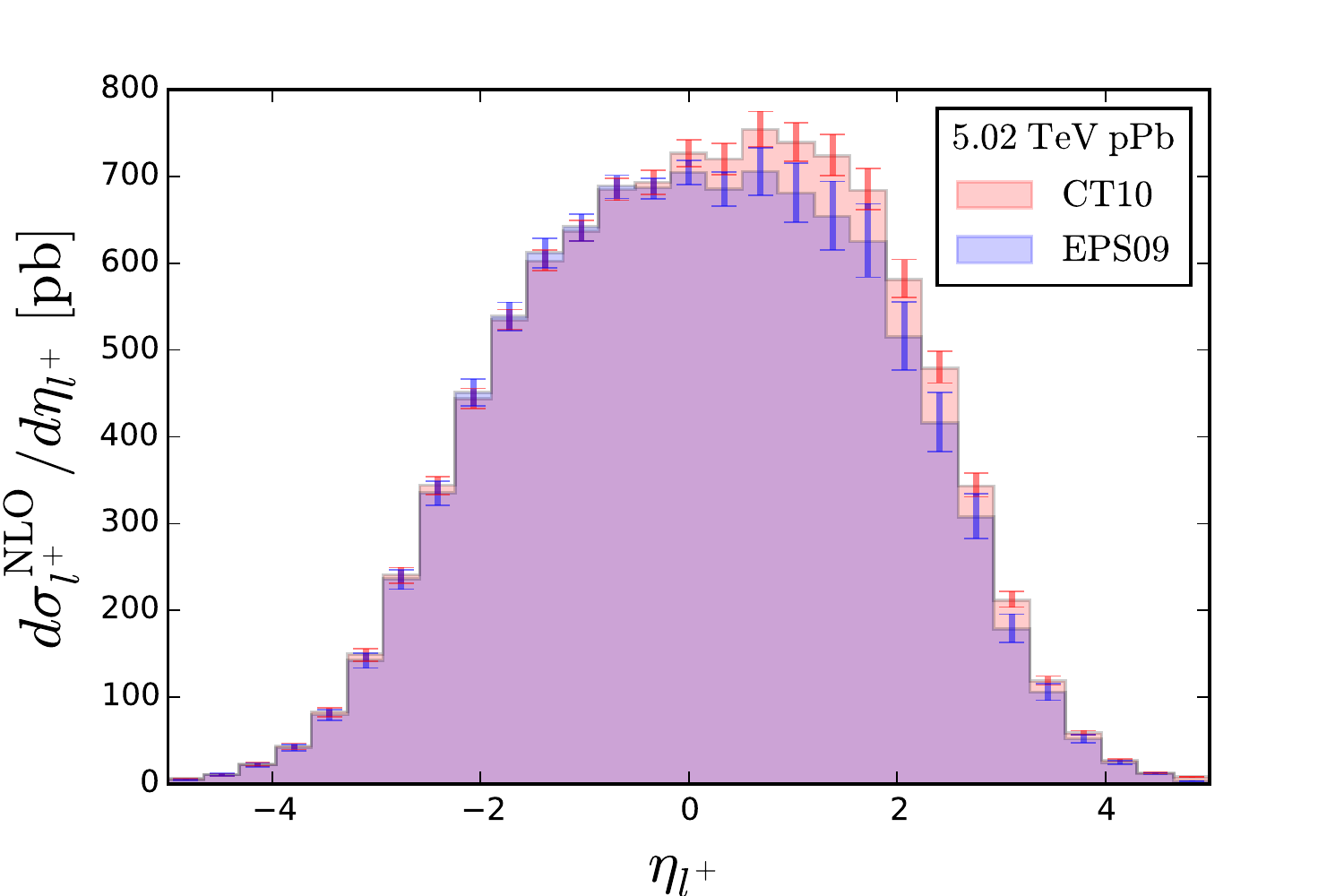}
\includegraphics[width=0.45\textwidth]{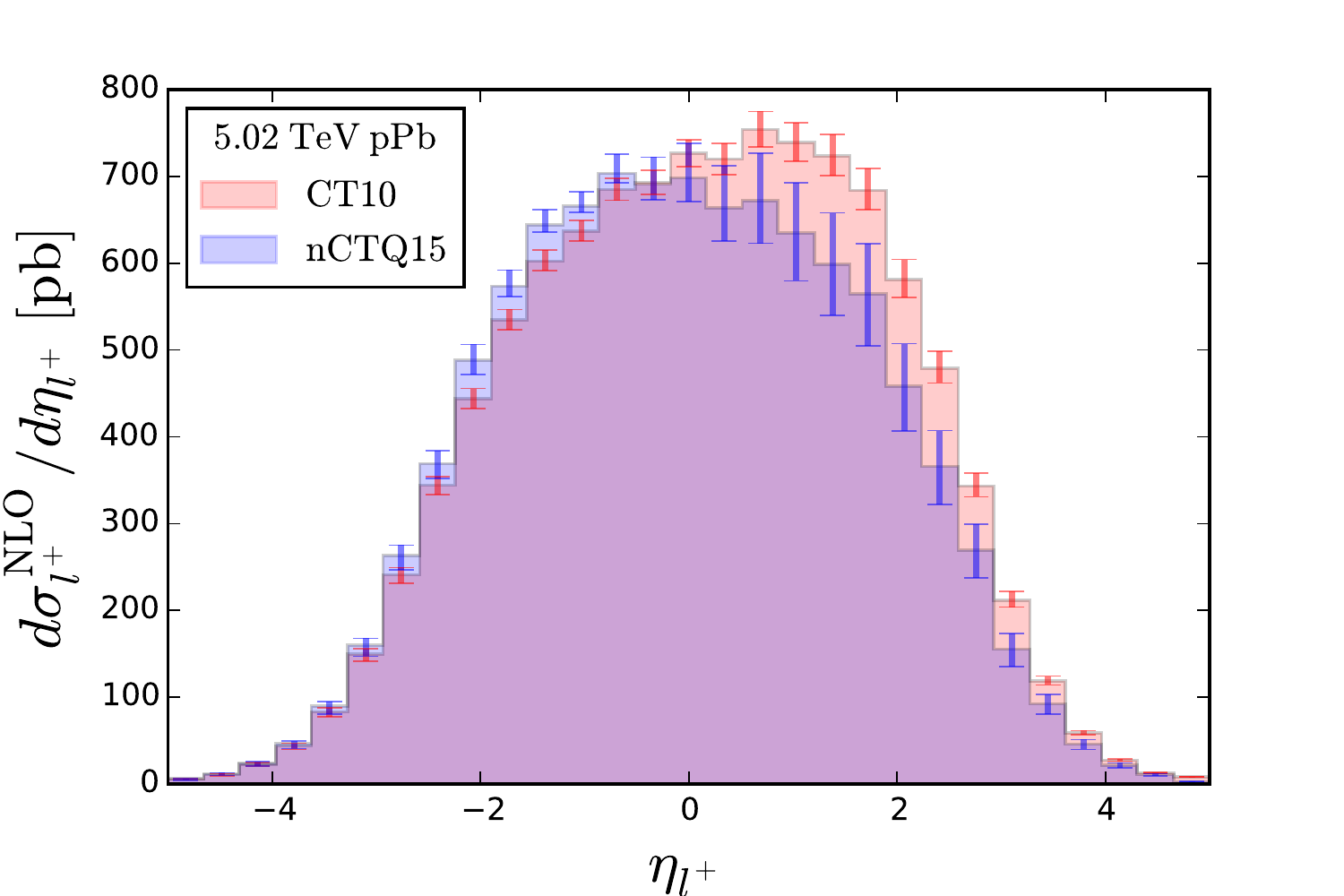}
\caption{The lepton rapidity distribution for pPb collisions at 5020 GeV at the LHC for EPS09 + CT10 
(left) and \texttt{nCTEQ15}~ (right) nPDFs compared to the distribution with 
the lead nucleus constructed 
from the CT10 free proton PDF. The initial state modifications to the PDF have a 
dramatic effect on the distribution.} 
\label{fig:lp_pPb}
\end{figure}

The Vector Boson cross section in pPb collisions at 5020 GeV provide a chance 
to see the nuclear modifications and compare predictions  
from different nPDF releases. In Fig.~\ref{fig:lp_pPb} we see a comparison of 
the differential cross sections calculated with \texttt{nCTEQ15}, 
EPS09 + CT10~\cite{Eskola:2009uj}, 
and CT10~\cite{Lai:2010vv}~PDFs. These distributions are asymmetric because of
different number of nucleons in each of the colliding nuclei.

The shape of the distribution can be understood by considering the LO 
calculation. In light-cone coordinates, the momentum fraction can be written 
$x = \tau e^{\pm y}$ where $+y$ is the rapidity of the beam with $x_1$, 
$-y$ is the rapidity of the beam with $x_2$, and $\tau = Q / \sqrt{S}$. For 
5020 GeV, we have $\tau \approx 0.016$ for on-shell $W^\pm$ 
production. This is equal to the momentum fraction at central rapidity. 
Moving to positive or negative rapidities is equivalent to a scan in the 
$x_1\ x_2$ plane. That is, as $y$ increases, $x_1$ increases and $x_2$
decreases. Therefore, scanning in $y$ probes the contributions to the PDF due 
to the nuclear modifications.

The asymmetry of the pPb cross sections can be anticipated by comparing the free 
proton PDF to the bound proton PDF in Fig.~\ref{fig:up_pdf}. At LO, the PDF can 
be thought of as the number density for each quark. At negative rapidities, the 
momentum fraction $x_2$ for the bound lead PDF increases and from 
Fig.~\ref{fig:up_pdf} we see that we move into an anti-shadowing region where the Pb 
PDF is increasing and the proton PDF is suppressed. The opposite occurs for 
positive rapidity. So we expect an enhanced (suppressed) number density in the 
region where $y < 0$ ($y > 0$). This corresponds to more (fewer) valance 
quarks available for the interaction.

The initial state modifications to the nPDFs have the startling effect of 
shifting the maximum of the cross section from the region of positive 
rapidity to the region of negative rapidity as seen in Fig.~\ref{fig:lp_pPb}. 
The combination of proton and neutron PDFs in the CT10 calculation 
enhances the cross section when $x_1$ is large and $x_2$ is small. 

We also compare the shapes of the distributions from \texttt{nCTEQ15} and EPS09 + CT10 
and find significant differences as seen in Fig.~\ref{fig:lp_pPb}. 
The differences are most pronounced in the high absolute rapidity region where 
the ratios of $u(x,Q) / d(x,Q)$ and $\bar{u}(x,Q) / \bar{d}(x,Q)$ become 
significant. Note that these ratios are one of the main differences between the various 
nPDF releases.   


\section{Conclusions}

Lepton rapidity distributions for on-shell vector boson production in PbPb and 
pPb collisions at 2760 and 5020 GeV at the LHC have been computed with 
\texttt{FEWZ\ 2.1}. The effects of the nuclear 
modifications to the PDFs are seen in the produced lepton rapidity distributions. 
In the PbPb case, the cross sections with nuclear modifications are consistent with a 
construction of the nucleus with free proton PDFs. 

Vector Boson production in pPb interactions provide an opportunity to observe 
the initial state modifications to the cross sections. The asymmetry inherent 
in the cross sections is heavily modified by the nuclear corrections. Additionally, 
these observables allow for comparison of different nPDF releases as differences 
in parameterization between them lead to quantitatively different results. 

The inclusion of LHC measurements in future nPDF fits will allow for stronger 
constraints on the PDF uncertainties.  



\addcontentsline{toc}{section}{References}
\bibliography{eprint_dpf2015.bbl}
\bibliographystyle{utphys}


\end{document}